# Thermal Aware Design Method for VCSEL-based On-Chip Optical Interconnect

Hui Li[1], Alain Fourmigue[2], Sébastien Le Beux[1*], Xavier Letartre[1], Ian O'Connor[1] and Gabriela Nicolescu[2]

[1] Lyon Institute of Nanotechnology, INL-UMR5270
Ecole Centrale de Lyon, Ecully, F-69134, France

[2] Computer and Software Engineering Dept.
Ecole Polytechnique de Montréal, Montréal (QC), Canada

* Contact author: sebastien.le-beux@ec-lyon.fr

*Abstract*— Optical Network-on-Chip (ONoC) is an emerging technology considered as one of the key solutions for future generation on-chip interconnects. However, silicon photonic devices in ONoC are highly sensitive to temperature variation, which leads to a lower efficiency of Vertical-Cavity Surface-Emitting Lasers (VCSELs), a resonant wavelength shift of Microring Resonators (MR), and results in a lower Signal to Noise Ratio (SNR). In this paper, we propose a methodology enabling thermal-aware design for optical interconnects relying on CMOS-compatible VCSEL. Thermal simulations allow designing ONoC interfaces with low gradient temperature and analytical models allow evaluating the SNR.

## I. INTRODUCTION

Technology scaling down to the ultra-deep submicron domain provides for billions of transistors on chip, enabling the integration of hundreds of cores. Many-core designs, integrating interconnect that can support low latency and high data bandwidth, are increasingly required in modern embedded systems to address the increasingly stringent power and performance constraints of embedded applications. Designing such systems using traditional electrical interconnect presents a significant challenge: due to capacitive and inductive coupling [5], both interconnect noise and propagation delay of global interconnect increase. The increase in propagation delay requires global interconnect to be clocked at a low rate, which limits the achievable bandwidth and system performance.

In this context, Optical Network-on-Chip (ONoC) is an emerging technology considered as one of the key solutions for future generations of on-chip interconnects. It relies on optical waveguides to carry optical signals, so as to replace electrical interconnect, and provide the low latency and high bandwidth characteristic of optical interconnect. Among the proposed ONoCs, the wavelength-routing based interconnect solutions are of considerable interest among the major players in the field, since they do not require any arbitration [1][2][4] to propagate the optical signals. They rely on passive Microring Resonators (MRs) that filter the optical signals based on their resonant wavelengths. However, silicon photonic devices are highly sensitive to temperature variation, which leads to a drift of the MRs resonant wavelength, and consequently a lower Signal to Noise Ratio (SNR). Furthermore, the power efficiency of integrated laser sources decreases at higher temperatures, which further decreases the interconnect power efficiency. Among the available laser sources candidates, Vertical-Cavity Surface-Emitting Laser (VCSEL) compatible to CMOS are of high of interest despite the use of less mature technologies (they usually require the inclusion of III–V semiconductors). Indeed CMOS-compatible VCSELs allow direct modulation of the optical signals and can thus be *dedicated* to a communication channel. Their size is of the same order of magnitude as the size of a MR used to modulate continuous waves emitted by *shared* lasers. VCSELs are thus sufficiently compact to be implemented in a large number and at any position, which leads to the following key advantages. First, the integration is easier since layout constraints are relaxed. Second, power saving is expected since waveguide lengths are reduced and waveguide crossings are avoided. Third, higher scalability is obtained since the laser sources are fully distributed.

For the first time, this paper proposes a thermal aware methodology to design ONoC relying on fully distributed CMOS-compatible VCSELs. The methodology relies on i) steady-state thermal simulations and ii) SNR analyses taking into consideration the temperature of the VCSEL and the MRs. Design space exploration on MR heating power and laser current modulation allow reducing ONoC gradient temperature while keeping acceptable the VCSEL temperature. SNR analyses allows estimating the ONoC reliability and power efficiency under a given chip activity.

The paper is structured as follows. Section II presents the related work. Then, the considered architecture models and laser sources are described in Section III. Section IV details the design methodology and Section V presents the case study and gives results. Section VI concludes the paper.

## II. RELATED WORK

Reducing the thermal sensitivity of optical communication has been addressed at both device and system levels. At device level, solutions relying on athermal devices [9], voltage tuning [10], local heating [11], and feed-back control schemes [12] have been explored to limit the thermal impact on or control the resonant wavelength of MRs. In [13], system level analyses allow evaluating the influence of temperature variation on the optical signal power received by the photodetectors. In order to counter-balance the temperature variation, communication channels can be remapped through ONoC reconfiguration [15] and DVFS and workload migration techniques can be applied [16]. These techniques depend on redundant resources to re-map the wavelength channel or detect run-time temperature variation.

In [14], the authors proposed a thermally-aware job allocation policy to minimize the temperature gradients among MRs. This work relies on device characteristics such as the Free Spectral Range (FSR), the number of waveguides and wavelengths. In our work, we evaluate the influence of the temperature variation on the SNR, taking into account the actual efficiency of the VCSEL under a given chip activity. In [21], the authors explore the placement of shared on-chip lasers on a layer located on top of the optical interconnect. In our work, we focus on CMOS-compatible VCSEL (allowing dedicated communication channel assignment) distributed on a layer dedicated to optical interconnects.

## III. 3D ARCHITECTURE

This section presents the architecture including ONoC. The challenges addressed in this paper are then introduced.

### A. Architecture Overview

Figure 1-a illustrates the MPSoC architecture we consider. It is composed of i) an electrical layer implementing processors (in tiles) and memories and ii) an optical layer dedicated to the implementation of a ring-based ONoC. The activity of the processors leads to local and global communications that are implemented with electrical interconnect and ONoC respectively. The communication hierarchy is defined at design time and depends mainly on the number of processors and the ONoC complexity and bandwidth. The silicon photonic fabrication process is compatible with the CMOS one allows integrating VCSEL, waveguides, MRs and photodetectors. These devices are gathered into so-called Optical Network Interface (ONI), which are responsible for emitting the light, modulating optical signals with the data to be transmitted, transporting the modulated signals and receiving them on the destination side (Figure 1-c). VCSEL and photodetectors are respectively connected to CMOS drivers and CMOS receiver through TSVs (Figure 1-c).

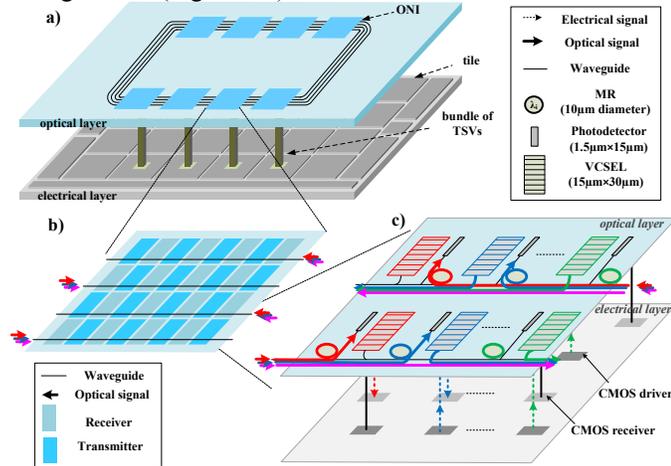

**Figure 1: Considered 3D architecture: a) MPSoC with a stacked ONoC, b) ONI layout and c) implementation of an ONI and its operations.**

We choose ORNoC [2] for the implementation of the optical communications. ORNoC is a ring-based network allowing point-to-point communications between source and destination, with passive MRs. As reported in [20], ORNoC demonstrates reduced worst-case and average insertion losses compared with related optical crossbars including Matrix [18], λ-router [1] and Snake [4] (e.g., on average, 42.5% reduction for worst-case and 38% for average in 4x4 scale), which is a significant advantage to reduce the laser power consumption.

### B. ONoC Interface and Thermal Sensitivity

The ONIs are responsible for emitting, transmitting and receiving the optical signals on the optical layer, as illustrated in Figure 1-c. For this purpose, the signals are injected into waveguides and modulated with VCSELs. The signals then propagate on the waveguide, crossing the intermediate ONIs. Once they reach their destination, the signals are dropped from the waveguide with passive MRs toward large band photodetectors. For proper filtering operation, it is mandatory that resonant wavelengths of destination MRs are well aligned with the wavelength of the emitted signals. However, silicon based optical devices are sensitive to temperature variation (0.1nm/°C typically), which leads to a reduced SNR due to reduced coupling of signals into MRs (i.e. lower signal power and higher crosstalk noise on photodetector).

Device-level calibration processes [16] help improving the SNR by aligning the resonant wavelength. However, such technique suffers from significant power consumption overhead: voltage tuning and heat tuning of MRs (that allow blue shift and red shift of the resonant wavelength respectively) lead to 130uW/nm and 190uW/nm respectively, as reported in [17]. For large-scale ONoCs (e.g. Corona [17] including approximately $1.1 \times 10^6$ MRs), the power dedicated to the calibration process represents more than 50% of the total network power. Since the calibration comes with performances overhead due to algorithm execution and heating latency, they are generally coupled to MRs clustering techniques. Indeed, clustering the MRs helps reducing the algorithm complexity by assuming a same local temperature among MRs close enough. However, this technique requires careful design of the ONIs to ensure a homogeneous temperature under different processing activities.

Maintaining low gradient temperature within an ONI including on-chip lasers is a challenging task since they dissipate a relatively high power. Heating MRs to reduce the gradient temperature is thus mandatory, which can be obtained by using a resistance on top of each MR. In addition, alternatively placing VCSEL and MRs contributes to reduce MRs heating power through a better initial distribution of the heat generated by VCSELs. This assumption leads to the chessboard-like layout illustrated in Figure 1-b: 4 waveguides propagating signals in clockwise and counter-clockwise rotation are alternatively placed and, for each waveguide, 4 receivers and 4 transmitters are alternatively placed.

### C. CMOS-Compatible VCSEL

VCSEL-based lasers [7][8] offer a direct emission of data through current modulation. While the fabrication processes of CMOS-compatible VCSEL is less mature than those of microdisk lasers [19], they offer significant advantage in terms of scalability (higher laser output power is achievable) and spectral density due to their small 3dB bandwidth (typically 0.1nm). The drawback of on-chip lasers over off-chip counterpart is their intrinsically lower efficiency and their higher sensitivity to the chip activity variation since they are located above the processing layer. More precisely, each VCSEL is located above a CMOS driver that converts an input electrical data coming from an IP core (represented as a binary voltage) into a current, as illustrated in Figure 2-a. The current propagates through a TSV and directly modulates the VCSEL. An optical signal is vertically emitted and is redirected to a horizontal waveguide through a taper. The power of the optical signal injected into the network ($OP_{net}$) thus depends on i) the intensity of the modulation current $I_{VCSEL}$, ii) the laser efficiency ($\eta_{laser}$) and the taper coupling efficiency ($\eta_{coupling}$, assumed to be 70%). The VCSEL efficiency is highly sensitive to its temperature: it can drop from 15% at 40°C to 4% at 60°C. This rather low efficiency leads to a high dissipated power ($P_{VCSEL}$) which, together with the power dissipated by the CMOS driver ($P_{driver}$) and chip ($P_{chip}$), influences the on-chip laser temperature. Hence, for a given modulation current, the power of the emitted optical signal ($OP_{VCSEL}$) depends on the laser temperature, which is influenced by $P_{chip}$, $P_{VCSEL}$ and $P_{driver}$, as illustrated in Figure 2-b. Furthermore, for an increasing activity of the processing layer (which is expected to result in additional communications), either the optical interconnect bandwidth will decrease assuming a same modulation current (the SNR being lower, data will be re-emitted) or the optical interconnect power consumption will increase (a higher modulation current is required to compensate the reduced efficiency). The modulation current thus must be carefully selected since i) a too small value will lead to low SNR and ii) a too high value will lead to a power hungry solution.

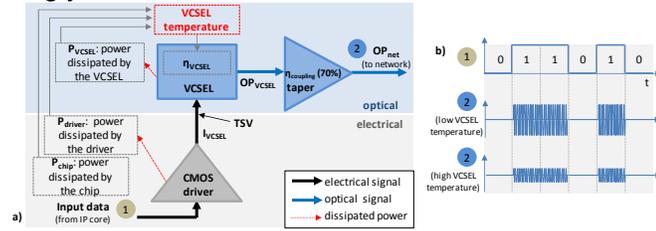

**Figure 2: The efficiency of a VCSEL depends on its temperature, which is influenced by the CMOS driver and the chip activity.**

*D. Contribution*

The gradient temperature and average temperature in ONI are critical to design VCSEL-based optical interconnect: i) **low gradient temperature** within an ONI eases the run-time calibration process and reduces the design complexity; ii) **low ONI average temperature** helps maintaining a reasonable power efficiency of the VCSEL.

For the first time, we propose a temperature aware design methodology allowing efficient use of CMOS-compatible VCSEL. Temperature evaluation and system level analyses allow estimating the ONoC SNR.

## IV. PROPOSED DESIGN METHODOLOGY

In this section, we describe the methodology allowing to design thermal-aware optical interconnect.

*A. Design Methodology Overview*

Figure 3 illustrates the design methodology. System level inputs include information on the packaging (heat sink base and fins, fan, etc.) and the considered architecture (die size, location of the ONIs, distance between the optical layer and the electrical layer, material used for each layer, etc.). The ONIs are more accurately described since it is possible to specify the number and the type of each device (VCSELs, MRs/heater, TSVs and photodetectors), their size and their relative location. Key parameters such as the $I_{VCSEL}$ and the MR heater power ($P_{heater}$) are specified by the user. The VCSEL electrical characteristics are fetched from a library and different activities can be considered to simulate the power dissipated by the processing layer (e.g. uniform and diagonal).

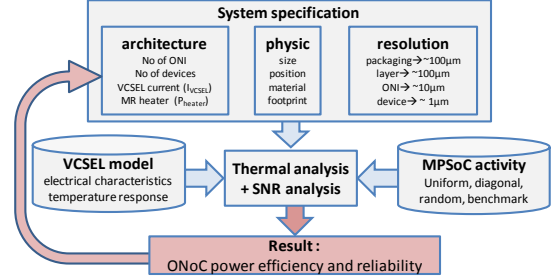

**Figure 3: Proposed methodology.**

Steady-state thermal simulations are achieved under a given chip activity. The result is a thermal map giving the gradient temperature and the average temperatures of each ONI. Reduction of the gradient temperature is obtained by exploring $P_{heater}$ values. SNR analyses are achieved to evaluate the optical interconnect reliability by considering the actual temperature of each ONI. The design space can be further explored by considering different $I_{VCSEL}$ values.

*B. System Specification*

In order to perform thermal evaluations, our architecture model is based on the real physical structure of the system. The different components of the system (i.e. package, die, heat sources, and optical devices) are represented as rectangular blocks, defined by their dimension, their position, and a constitutive material. The blocks can be assigned to power values, which allow modeling the heat sources of the system. The Back-End-Of-Line (BEOL) is modeled as a thin layer (10µm) and the heat sources (cores, cache, router, etc.) are represented as rectangular blocks with power values, situated in the BEOL layer.

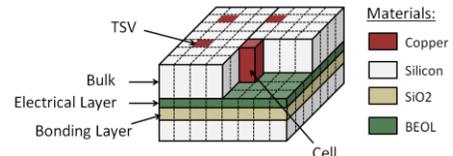

**Figure 4: IcTherm computes the heat transfers between the cells and outputs the temperature value of each cell. This thermal map allows computing the gradient temperature between any points of the system.**

$IcTherm^1$ [23] is a thermal simulator for electronic devices which accurately models their complex structure and provides 3D full-chip temperature maps. IcTherm solves the physical equations that govern the temperature in the chip, using the Finite Volume Method [24], a numerical method for solving partial differential equations. IcTherm was validated against the commercial simulator COMSOL [22]: its maximal error was found to be less than 1% [23]. The structure of the system is discretized into small cubic cells that match the distribution of the materials and the heat sources. Figure 4 illustrates the discretization of a section of the system. Because the interfaces contain micro-scale components (e.g. TSVs, VCSELs and CMOS drivers), we use a fine-grain resolution with a cell size of 5 μm x 5 μm for meshing the region containing the interfaces. For the rest of the system, we use a coarser resolution with a cell size of 100 μm x 100 μm for the heat sources and 500 μm x 500 μm for the package.

---

[1] IcTherm website : http://www.ictherm.com/

## C. SNR Analysis

For a given activity scenario, the thermal map gives the temperature of the lasers and the MRs, from which the gradient temperature of each ONI is extracted. We assume that the gradient temperature within an ONI must remain below 1°C (i.e. since we consider MRs with 1.55nm 3dB bandwidth, 0.1nm drift of their resonant wavelength corresponds to 6.5% transmission loss). Design space on the MRs heater power can be explored in order to satisfy the 1°C gradient temperature constraint *inside* each ONI. However, the temperature gradient *among* ONIs influences the SNR, as detailed in the following.

Figure 5-a illustrates the transmission of an optical signal $OP_{in}$ at wavelength $\lambda_{signal}$ into an MR with a resonant wavelength $\lambda_{MR}$. The 3dB bandwidth of the signal is assumed to be small compared to MR's one (0.1nm and 1.55nm respectively). The level of the signal power at the through port ($OP_{through}$) and drop port ($OP_{drop}$) depends on the alignment between $\lambda_{signal}$ and $\lambda_{MR}$, (Figure 5-b). A maximum transmission to the drop port is obtained when $\lambda_{signal}=\lambda_{MR}$. In case both wavelengths are significantly different (above 1.5nm), most of the input signal power continue propagating along the waveguide to the through port. The misalignment of the wavelengths may be a required from proper ONoC operation (for wavelength routed routing) or it can be a side effect related to a temperature difference between the ONIs. By assuming $\lambda_{signal}$ equals to $\lambda_{MR}$ for a same laser source and MR temperature, 50% of the signal will be (wrongly) dropped from the waveguide for a 7.7°C temperature difference (0.77nm misalignment). This will lead to crosstalk and signal attenuation, which is evaluated with an analytical model.

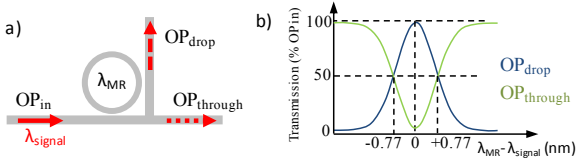

**Figure 5 MR transmission model: a) MR structure, b) signal transmission according to the alignment between $\lambda_{signal}$ and $\lambda_{MR}$.**

Figure 6 illustrates a waveguide propagating optical signals injected and ejected by the ONIs. Since we consider ORNoC [2], no arbitration is needed and passive MRs are used (i.e. the MRs resonant wavelength is defined at design time but it shifts with the temperature variation). A communication between a source interface $ONI_S$ and a destination interface $ONI_D$ (i.e. denoted as $C_{s \to d}$ and represented by the red line) implies that an optical signal is 1) generated and modulated by a VCSEL ($T_{s \to d}$) and 2) dropped to a photodetector by a MR ($R_{s \to d}$). The quality of the communication $C_{s \to d}$ depends upon the insertion loss along the path and the crosstalk induced by the other communications $C_{i \to j}$, where i≠s and j≠d (e.g. the blue line). The worst-case SNR for $C_{s \to d}$ is formalized as follow:

$$SNR = 10\log_{10}\frac{P_{signal}}{P_{noise}} = 10\log_{10}\frac{OP_{s-d}[s \to d]}{\sum_{i=1}^{N}\sum_{j=1}^{N}X_{s-d}[i \to j]}$$

with i≠j and i→j ≠s→d,

Where $OP_{s \to d}[s \to d]$ is the expected power of signal $C_{s \to d}$ received by $R_{s \to d}$ (in $ONI_D$), and $X_{s \to d}[i \to j]$ is the crosstalk power of the communications $C_{i \to j}$ received by $R_{s \to d}$.

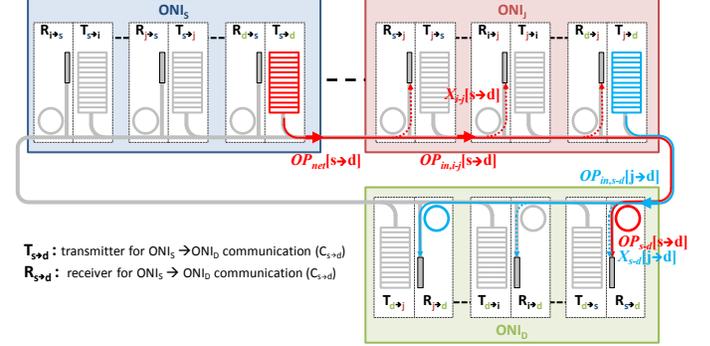

**Figure 6: Signal propagation in a waveguide. The quality of a communication depends on the insertion losses and the crosstalk.**

The losses encountered by the signal $C_{s \to d}$ depend on the waveguide length and the crossed MRs (in $R_{i \to j}$) in the intermediate $ONI_J$. In case signal wavelength $\lambda$ of $C_{s \to d}$ (denoted $\lambda S_{s \to d}$) is far enough from the MR resonant wavelength $\lambda R_{i \to j}$ (in $R_{i \to j}$), the losses are expected to be small. This optimistic scenario occurs if the temperature of the intermediate ONI ($ONI_J$ in the figure) is strictly equal to the temperature of the source ONI. A temperature difference among different ONIs will lead to a misalignment of the wavelengths: part of the signal power will be ejected from the waveguide and will reach a wrong photodetector. This leads to additional crosstalk and reduced signal power on intermediate and destination photodetectors respectively (e.g. $X_{s-d}[j \to d]$ and $OP_{s-d}[s \to d]$ in the figure). The losses are formulated as follows:

$$X_{i \to j}[s \to d]=OP_{in,i \to j}[s \to d]\times\Delta\lambda_{i \to j}[s \to d]$$

$$OP_{in,i-j}[s \to d] = OP_{net}[s \to d]\times \prod_{k=s}^{\substack{j-1,if\ j>s \\ j+N-1,if\ j<s}} L_{k \bmod N} \times$$

$$\prod_{m=s+1}^{\substack{j,if\ j>s \\ j+N,if\ j<s}} \prod_{k=1}^{\substack{i-1,k\neq m\bmod N,if\ m=j \\ N,k\neq m\bmod N,if\ m\neq j}} (1-\Delta\lambda_{k-m \bmod N}[s \to d])$$

$$L_k = L_{propagation}^{l_k}$$

Where $X_{i \to j}[s \to d]$ is the signal power dropped to $R_{i \to j}$, $OP_{in,i \to j}[s \to d]$ is the power of the signal at the $R_{i \to j}$, $\Delta\lambda_{i \to j}[s \to d]$ is the signal ratio dropped by MR (in $R_{i \to j}$, $\Delta\lambda_{k \to 0}[s \to d]=\Delta\lambda_{k \to N}[s \to d]$). $N$ is the number of ONIs in the optical interconnect. $L_k$ is the propagation loss (in %) along a communication path (e.g. $L_1$ for $C_{1 \to 2}$ and $L_2$ for $C_{2 \to 3}$), $l_k$ is the corresponding length of the waveguide and $L_{propagation}$ is the propagation loss (in % per cm).

The crosstalk noise received by each photodetector is estimated by considering the total signal power dropped due to wavelengths misalignment and is formulated as follows:

$$X_{s \to d}[j \to d]=OP_{in,s \to d}[j \to d]\times \Delta\lambda_{s-d}[j \to d]$$

SNR analysis allow estimating the ONoC reliability under a given chip activity. This crucial information allows the exploration of the design space and particularly the driver power consumption. Indeed, $P_{driver}$ is directly related to the laser modulation current and, therefore, it impacts the laser efficiency and the optical signal power. Such exploration is illustrated in the following section.

## V. RESULTS

We first detail the considered architecture and VCSEL model. Then, MR heater power is explored through thermal simulations in order to reduce the gradient temperature. Finally, ONoC solutions are compared according to the SNR.

### A. Case Study

The targeted system used in our experiments is based on Intel's Single-Chip Cloud Computer (SCC) [6], a 24-tile, 48-core IA-32 45nm processor with a maximum power dissipation of 125W. Given the large number of cores, the SCC is a good candidate for silicon-photonic communications. We model the same package as the one used by Intel [6]. Figure 7 shows the assembly view of the targeted system, which contains the following components: steel back-plate, motherboard, socket, SCC chip with silicon-photonic links and on-chip laser sources, copper lid and heat sink.

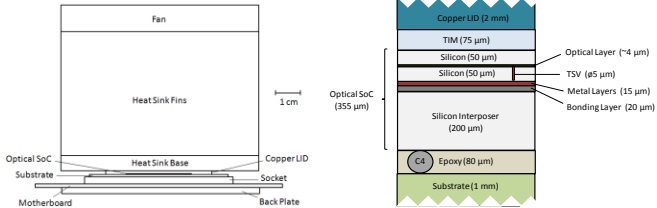

**Figure 7: Packaging of the SCC chip and the optical interconnect.**

We assume ORNoC relying on 4 waveguides and 4 lasers per waveguide per ONI. The CMOS drivers and receivers are placed on an empty area of the electrical router of SCC tile. We consider a VCSEL with a 15x30µm² footprint size [7][8]. It relies on mirrors redirecting the vertically generated light into to horizontal waveguide, which allow reducing the thickness of the laser below 4µm (i.e. thus making the VCSEL compatible to CMOS). The direct modulation bandwidth is 12GHz and the 3dB bandwidth is about 0.1nm.

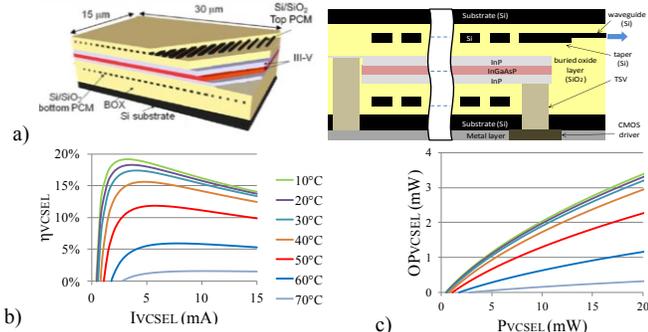

**Figure 8: VCSEL: a) 3D view extracted from [7], b) efficiency and c) laser output power according to the temperature.**

Figure 8-a represents a 3D view of the laser extracted from [7]: it is composed of 3 layers of III-V material (0.6µm, 0.45µm and 0.4µm thickness respectively). The laser effect is generated in the central active layer (in red color) and the power is dissipated in the adjacent layers (in grey color). Two contacts allow driving the current from the CMOS layer through 5µm-diameter TSVs. The active layers are surrounded by Si/SiO$_2$ line constituting the mirror structure, which allows coupling the light into the horizontal waveguide (optical signal shown in blue arrow) using a tapper with an estimated 70% efficiency. Figure 8-b gives the laser efficiency according to its temperature and $I_{VCSEL}$. Figure 8-c gives the influence of $P_{VCSEL}$ and its temperature on the actual emitted light $OP_{VCSEL}$.

### B. Reduction of the ONI Gradient Temperature

We first evaluate the influence of $P_{chip}$ and $P_{VCSEL}$ on the ONI average and gradient temperature. We run thermal simulations under homogeneous 12.5W, 18.75W, 25W and 31.25W chip activities and we explored $P_{VCSEL}$ ranging from 0 to 6mW (we assumed $P_{VCSEL}=P_{driver}$, which corresponds to the worst case scenario since the total energy received by the VCSEL is dissipated as heat). Figure 9-a illustrates the average temperature results: a 6W increase of the total chip activity roughly leads to +3.3°C on the average temperature while a 6mW increase of $P_{VCSEL}$ leads to +11°C. This demonstrates how important the calibration of the laser modulation current according to the requirements is: a slightly over-sized current will lead to significant power consumption overhead.

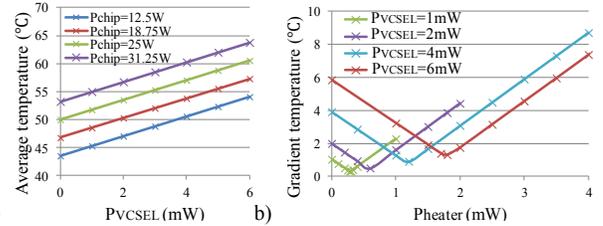

**Figure 9: a) Influence of $P_{VCSEL}$ and $P_{chip}$ on the average temperature and b) influence of $P_{VCSEL}$ and $P_{heater}$ on the gradient temperature.**

Results also show a significant impact of $P_{VCSEL}$ on the gradient temperature between lasers and MRs (1.7°C/mW). Such gradient temperature does not realistically allow using clustering technique for run-time calibration process. We thus explore MR heating power values ($P_{heater}$) to reduce this gradient at design time. As illustrated in Figure 9-b, the smallest gradient is obtained for $P_{heater}= 0.3 \times P_{VCSEL}$. In Figure 10, we compare the temperature results with and without MR heater: for $P_{VCSEL}=1$mW, solutions with and without heater lead to 0.3°C and 1°C gradient temperature respectively. Significant improvement of the gradient temperature is obtained for higher $P_{VCSEL}$ values: for 6mW, the gradient temperature drops from 5.8°C to 1.3°C (i.e. -4.5°C), which is significant compared to the reasonable 0.8°C increase of the average laser temperature.

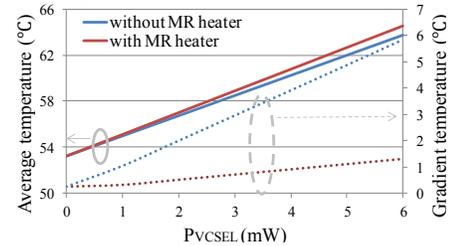

**Figure 10: Average and gradient temperatures w/ and w/o MR heater.**

## C. System Level Estimation of SNR

We evaluate the influence of the ONIs location on the SNR. As illustrated in Figure 11, we consider 3 scenarios leading to 18mm (in red), 32.4mm (in blue) and 46.8mm (in green) waveguide lengths. $P_{VCSEL}$ and $P_{heater}$ are set to 3.6mW and 1.08mW respectively (simulations validated that the gradient temperature within each ONI remains below 1°C). $OP_{VCSEL}$ is estimated from the ONI average temperature and Figure 8-c.

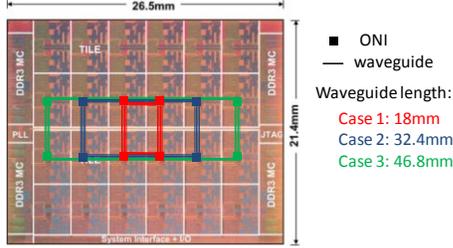

**Figure 11: The location of the ONI leads to the following waveguide lengths: a) 18mm, b) 32.4m and c) 46.8mm.**

**Table 1: Technological parameters**

| Parameters | Value |
|---|---|
| Wavelength range | 1550nm |
| $BW_{3\text{-}dB}$ | 1.55 nm |
| Photodetector sensitivity | -20dBm (0.01mW) |
| Thermal sensitivity | 0.1nm/°C |
| $L_{propagation}$ | 0.5 dB/cm [3] |

Table 1 summarizes the technological parameters we have considered and Figure 12 gives the worst-case SNR results. Under a uniform activity, the asymmetric structure of the SCC chip leads to a 3°C difference among the ONIs for 46.8mm length. The crosstalk is relatively small and the SNR thus only depends on the signal power, which depends on the propagation losses. The SNR drops from 38dB for 18mm length to 13dB for 46.8mm. Under the diagonal activity, the upper-right and bottom-left parts of the chip dissipate each 4W while the upper-left and bottom-right parts dissipate 8W each. This leads to heterogeneous temperature of the ONIs (54.62→55.92°C for case 1, 54.33→56.92°C for case 2, 56.16→60.85°C for case 3 respectively). Compared to the uniform activity, the diagonal activity exhibits lower average temperature since upper-right and bottom-left parts have lower power. This leads to higher laser efficiency, further resulting in higher $OP_{VCSEL}$. However, the temperature gradient *among* the ONIs being higher, additional crosstalk occurs, which results in a lower SNR compared to the uniform activity. A random activity leads to intermediate SNR results.

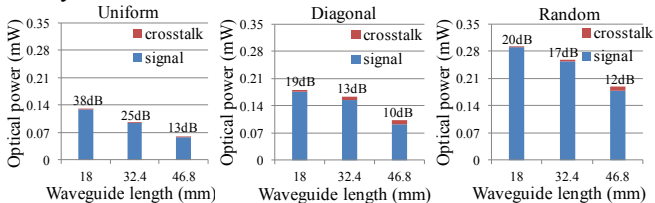

**Figure 12: SNR results under uniform, diagonal and random activities**

This analysis validates that the ONoC matches with the receiver sensitivity and SNR requirements. Further explorations of the design space allow optimizing the ONoC. For instance, in case a lower SNR is acceptable, $P_{VCSEL}$ and $P_{heater}$ can be reduced for energy saving.

## VI. CONCLUSION

In this paper, we proposed a thermal aware design methodology for CMOS-compatible VCSEL-based ONoC. Thermal simulations allow exploring design parameters such as the laser modulation current and heater power in order to reduce the gradient temperature within an ONoC interface. A heater dissipating 30% of the VCSEL power leads to an optimal solution for the considered architecture. SNR analyses allow comparing ONoC under different chip activities.

### ACKNOWLEDGMENT
Hui LI is supported by China Scholarship Council (CSC).

### REFERENCES

[1] I. O'Connor, et al. "Reduction Methods for Adapting Optical Network on Chip Topologies to Specific Routing Applications". In Proceedings of DCIS, 2008.

[2] S. Le Beux, et al. "Layout Guidelines for 3D Architectures including Optical Ring Network-on-Chip (ORNoC)". In 19th IFIP/IEEE VLSI-SOC 2011

[3] A. Biberman, et al. "Photonic Network-on-Chip Architectures Using Multilayer Deposited Silicon Materials for High-Performance Chip Multiprocessors", ACM Journal on Emerging Technologies in Computing Systems 7 (2) 7:1-7:25

[4] Luca Ramini, et al. "Contrasting wavelength-routed optical NoC topologies for power-efficient 3D-stacked multicore processors using physical-layer analysis". In Proceedings of DATE, 2013.

[5] R. Ho, et al. The Future of Wires. Proceedings of the IEEE, 89(4):490–504, 2001

[6] J. Howard et al. A 48-core IA-32 processor in 45 nm CMOS using on-die message-passing and DVFS for performance and power scaling. Solid-State Circuits, IEEE Journal of, 46(1):173–183,. 2011

[7] C. Sciancalepore, et al. Thermal, Modal, and Polarization Features of Double Photonic Crystal Vertical-Cavity Surface-Emitting Lasers. IEEE Photonics journal. Vol. 4, No 2, April 2012.

[8] Markus-Christian Amann and Werner Hofmann. InP-Based Long-Wavelength VCSELs and VCSEL Arrays. IEEE journal of Selected Topics in Quantum Electronics Vol. 15, No. 3, 2009.

[9] S. S. Djordjevic, et al. "CMOS-compatible, athermal silicon ring modulators clad with titanium dioxide", OPTICS EXPRESS, Vol. 21, No. 12, 2013.

[10] S. Manipatruni, et al., "Wide temperature range operation of micrometer-scale silicon electro-optic modulators", OPTCS LETTERS, Vol. 33, No. 19, , 2008.

[11] A. Biberman, et al., "Thermally Active 4x4 Non-Blocking Switch for Networks-on-Chip", in IEEE Lasers and Electro-Optics Society, pp. 370-371, 2008.

[12] K. Padmaraju, et al., "Thermal stabilization of a microring modulator using feedback control", Optics Express, Vol. 20, No. 27, pp. 27999-28008, 2012.

[13] Yaoyao Ye, et al, "System-Level Modeling and Analysis of Thermal Effects in Optical Networks-on-Chip", IEEE Transactions on Very Large Scale Integration Systems, vol. 21, no. 2, pp. 292-305, February 2013.

[14] T. Zhang, et al., "Thermal management of Manycore Systems with Silicon-Photonic Networks," Proc. Design, Automation and Test in Eutope (DATE) 2014.

[15] Y. Zhang, et al., "Power-Efficient Calibration and Reconfiguration for Optical Network-on-Chip", Journal of Optical Communications and Networking, 2012.

[16] Z. Li, et al, "Reliability Modeling and Management of Nanophotonic On-Chip Networks", IEEE Transactions on Very Large Scale Integration (VLSI) Systems, Vol. 20, No. 1, pp. 98-111, 2012.

[17] J. Ahn, et al. "Devices and architectures for photonic chip-scale integration". Applied Physics A: Materials Science & Processing, 95(4):989–997, 2009.

[18] A. Bianco, et al. "Optical Interconnection Networks based on Microring Resonators". In IEEE International Conference on Communications, 2010.

[19] J. Van Campenhout, et al., "Electrically pumped InP-based microdisk lasers integrated with a nanophotonic silicon-on-insulator waveguide circuit", Optics Express, 15(11), p.6744-6749, 2007.

[20] S. Le Beux, et al.. "Optical Crossbars on Chip, A Comparative Study based on Worst-Case Losses". In Wiley CCPE, 2014.

[21] C. Chen, et al, "Sharing and Placement of On-chip Laser Sources in Silicon-Photonic NoCs," Proc. International Symposium on Networks-on-Chip, 2014.

[22] COMSOL. http ://www.comsol.com, July 2014.

[23] A. Fourmigue, et al"Efficient Transient Thermal Simulation of 3D ICs with Liquid-Cooling and Through Silicon Vias", DATE 2014.

[24] S. C. Chapra and R. P. Canale. Numerical Methods for Engineers. McGraw-Hill, Inc., New York, NY, USA, 6th edition, 2009.